\documentclass[runningheads,a4paper]{llncs}

\usepackage{hyperref}
\usepackage{amssymb}
\usepackage{graphicx}
\usepackage{url}
\usepackage{apacite}
\usepackage{amsmath}
\usepackage{scrextend}
\usepackage{microtype}

\newcommand{\keywords}[1]{\par\addvspace\baselineskip\noindent\keywordname\enspace\ignorespaces#1}
\renewcommand{\cite}[1]{\shortcite{#1}}

\usepackage{calc}
\usepackage[absolute]{textpos}
\newlength{\footerxpos}
\newlength{\footerypos}
\newcommand{\copyrightnote}{\copyright\:2020 Copytight held by the owner/author(s). This is the author's version of the work. It is posted here for your personal use. Not for redistribution. The definitive Version of Record will be published in \textit{Proceedings of the 2020 Joint Conference on AI Music Creativity} (CSMC-MuMe 2020), ISBN 978-91-519-5560-5.}

\begin{document}

%
\begin{textblock*}{\columnwidth}(\footerxpos,\footerypos)\noindent\footnotesize{\copyrightnote}\end{textblock*}

\mainmatter

\title{Generative Melody Composition with Human-in-the-Loop Bayesian Optimization}

\titlerunning{Generative Melody Composition with Human-in-the-Loop BO}

\author{Yijun Zhou\inst{1}, Yuki Koyama\inst{2}, Masataka Goto\inst{2}, \and Takeo Igarashi\inst{1}\thanks{This work was supported in part by JST CREST Grant Number JPMJCR17A1, Japan, and JST ACCEL Grant Number JPMJAC1602, Japan.}}

\authorrunning{Yijun Zhou, Yuki Koyama, Masataka Goto, and Takeo Igarashi}

\institute{The University of Tokyo \and National Institute of Advanced Industrial Science and Technology (AIST)}

\maketitle

\begin{abstract}
  Deep generative models allow even novice composers to generate various melodies by sampling latent vectors.
  However, finding the desired melody is challenging since the latent space is unintuitive and high-dimensional.
  In this work, we present an interactive system that supports generative melody composition with human-in-the-loop \emph{Bayesian optimization} (BO).
  This system takes a mixed-initiative approach;
  the system generates candidate melodies to evaluate, and the user evaluates them and provides preferential feedback (i.e., picking the best melody among the candidates) to the system.
  This process is iteratively performed based on BO techniques until the user finds the desired melody.
  We conducted a pilot study using our prototype system, suggesting the potential of this approach.

\end{abstract}

\section{Introduction}

Deep generative models, such as generative adversarial networks (GANs) \cite{10.5555/2969033.2969125} and variational auto-encoders (VAEs) \cite{DBLP:journals/corr/KingmaW13}, have been applied to support creative work, showing great potentials in images, videos, and music \cite{10.5555/3157096.3157165, Zhu2016GenerativeVM, DBLP:conf/aaai/DongHYY18}.
When we focus our scope on the music domain, many research projects have tried to learn latent spaces of melodies \cite{YangCY17, 10.5555/2591248.2591260, Simes2019DeepLF, 10.1145/3108242}.
By sampling latent vectors from the smooth latent space produced by one of these generative melody models, even novice composers can quickly generate various meaningful melodies \cite{Ghosh2019FromVT, Roberts:ICML:18}.

However, end-users may face significant challenges when exploring the latent space to find a desired melody, since the latent space is often unintuitive and high-dimensional.
Researchers have developed many interaction techniques (e.g., mixing two melodies) to assist users in exploring the latent space to achieve their creative goals in music~\cite{10.1145/2876456.2879471, magentajs}.
These techniques have shown opportunities to enhance the user experience of music composition.

In this work, we propose a new approach to facilitate generative melody composition for novices, which is based on human-in-the-loop \emph{Bayesian optimization} (BO) techniques \cite{Brochu:NIPS:07, Koyama:SIGGRAPH:17, Koyama:SIGGRAPH:20}.
Specifically, we present a melody-composition system that employs a human-AI mixed-initiative interaction:
the system generates candidate melodies to evaluate, and the user evaluates them and provides preferential feedback (i.e., picking the best melody among the observed candidates) to the system.
This process is iteratively performed until the user finds the desired melody.
The generation of candidate melodies by the system is based on BO, which tries to make the query to the user as effective in terms of the number of necessary iterations as possible.

We evaluated our approach through both a simulated experiment and a pilot user study, suggesting the potential of this approach.
To the best of our knowledge, this work is the first to investigate the potential of human-in-the-loop BO in the context of generative melody composition and AI music creativity.

\section{Related Work}

\subsection{Human-in-the-Loop Bayesian Optimization}

BO is a black-box optimization technique \cite{Shahriari:ProcIEEE:16}.
It is especially useful when handling expensive-to-query objective functions since it tries to minimize the number of necessary iterations by selecting the most effective query in each iteration.
Thus, it is widely used for, for example, hyperparameter tuning of deep learning models \cite{Akiba:KDD:19}.

Human's preference is expensive to query.
Thus, a search problem with an objective function defined by human perceptual assessment (e.g., preference) should be solved with a minimal number of queries.
This motivates researchers to develop human-in-the-loop BO methods \cite{Brochu:NIPS:07, Koyama:SIGGRAPH:17, Koyama:SIGGRAPH:20, Chong2019InteractiveSE}.
As preferential assessment needs to be ``relative'' (e.g., which option is the better?) rather than ``absolute'' (e.g., how good is this option?) \cite{Brochu:SCA:10, Tsukida:TR:11}, these methods take relative preferential feedback as input, unlike typical BO algorithms.
Thus, this approach is also called as \emph{preferential BO} (PBO).
The PBO method proposed by Brochu et al.\ \citeyear{Brochu:NIPS:07} takes pairwise-comparison results as input, which is the simplest form of preferential feedback.
To make the necessary number of queries even smaller, the methods proposed by Koyama et al.\ \citeyear{Koyama:SIGGRAPH:17, Koyama:SIGGRAPH:20} take subspace-search results as input, where human is iteratively asked to find the best possible option from a one- or two-dimensional search subspaces.

In this work, we apply this PBO\footnote{In the following sections, we simply call it BO when no confusion can arise.} approach to generative melody composition.
Our system iteratively asks the user to compare multiple melody candidates sampled from a one-dimensional subspace \cite{Koyama:SIGGRAPH:17} and to select the preferred one.
In addition, we allow the user to directly edit the selected one to provide more controllability to the user.

\subsection{Tools for Deep Music Generation}

The expanding use of deep generative models encourages researchers to work on interfaces and tools to make the interactive music generation experience more accessible \cite{46821, 48280}.
Vogl et al.\ \citeyear{10.1145/2876456.2879471} presented an AI-powered software prototype for generating drum pattern variations.
Esling et al.\ \citeyear{FlowSynthesizer} applied a VAE to increase the controllability of sound synthesizers.
Roberts et al.\ \citeyear{Roberts:ICML:18} proposed a recurrent VAE, \emph{MusicVAE}, which utilized a hierarchical decoder for improved modeling sequences with long-term structure.
Our system uses the pre-trained MusicVAE, considering its smooth latent space.
\emph{Magenta.js} \cite{magentajs} provides developers with APIs to design and deploy web-based interface of generative melody models.
Louie et al.\ \citeyear{Louie:CHI:20} investigated \emph{AI-steering} tools to give users more control over exploring latent spaces of deep generative models, and presented a human-AI co-creation tool.
While we share a similar motivation, we focus on investigating the approach of formulating the task as a human-in-the-loop optimization.

\section{System Overview}

\begin{figure}
  \centering
  \includegraphics[width=\textwidth]{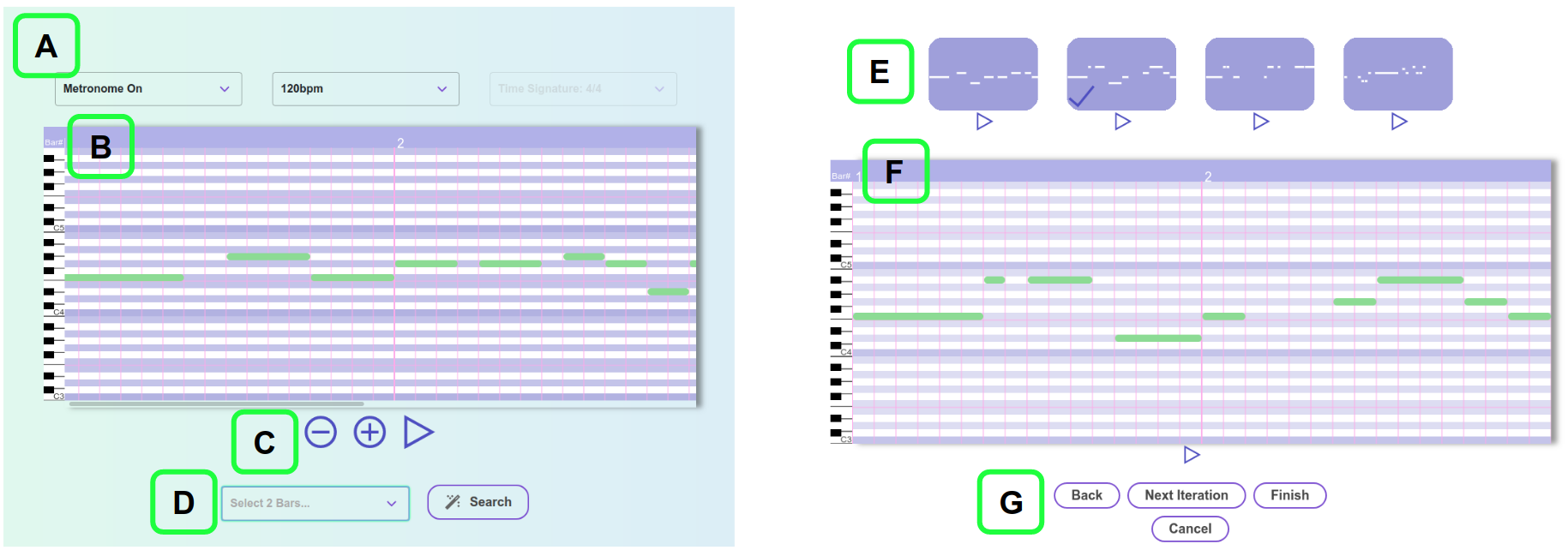}
  \caption{
    System interface. (Left) Composing mode: (A) metronome on/off and BPM; (B) note sheet; (C) add/subtract more bars and play; (D) select two bars and open the searching mode. (Right) Searching mode: (E) current candidates that BO generated; (F) editing widget; (G) back to the initial state, go to the next iteration, finish, or cancel the searching.
  }
  \label{fig:system}
\end{figure}

Our system is web-based, and it consists of two main modes: composing mode (\autoref{fig:system} (left)) and searching mode (\autoref{fig:system} (right)).
The searching mode works as a dialog of the system.
The front-end is implemented in JavaScript, while the back-end is managed by a Flask (Python) server \cite{10.5555/2621997}.
All the musical data represented on the interfaces and transferred between front-end and back-end are on the symbolic MIDI representation \cite{10.5555/525217}.
A video demo of the system can be found at our web source\footnote{\label{note1}\href{http://yijunzhou.xyz/index/generative-melody-composition-with-human-in-the-loop-bo}{yijunzhou.xyz/index/generative-melody-composition-with-human-in-the-loop-bo}}.

\paragraph{Composing Mode.}
The composing mode allows the user to compose melodies by inputting notes.
The interface resembles standard composing software interface, providing a piano-roll timeline and essential functions.
Vertical axis represents pitch, and horizontal axis
represents timeline.
Each column represents a 16th note duration, and
a set of 16 columns represents a bar.
The user can change the length of the melody by
adding and removing bars.
Each row represents a pitch of piano key.
Since we focus on the main melody composition, we limit the pitch range into three octaves, from C3 to B5.
The user can draw and delete notes, and change notes' pitches on the note sheet.
The composed melody is used as the starting point of the search when entering the searching mode.

\paragraph{Searching Mode.}
The searching mode is designed to iteratively show multiple candidates generated by BO based on the user's preference.
In each iteration, it shows $n$ candidates ($n = 4$ in our case), from which it allows the user to select the one they prefer.
After selecting a candidate, they can edit it in the editing widget before going to the next iteration.
The user is not allowed to edit the other bars from the composing mode in the editing widget.
They can always go back to the initial state of the current iteration.
The iteration goes on by showing the user another batch of candidates until the user finds the desired one.

\paragraph{Interaction Procedure.}

(1) Every time the user gets stuck at composing, they can select a two-bar melody clip that they wish to see variations and then click the ``Search'' button in the composing mode.
(2) After the searching mode pops up, the user will see the original melody clip in the editing widget and $n$ candidates on the top, which are drawn by BO from the latent space\footnote{At the first iteration, we generate candidates by random sampling since BO does not have data yet.};
the user can listen to them and select the most favorite one among the candidates and the current melody.
(3) The candidate the user selected will be shown in the editing widget, and the user can tweak details if they want.
(4) After editing the candidate, the user can click the ``Next Iteration'' button to go to the next iteration, where they will see a new batch of candidates generated by BO based on their preferential input;
the edited melody in the last iteration stays at the editing widget.
(5) The user can go back to the initial state of the current iteration by clicking ``Back''.
(6) Once the user finds the desired one, they can click the ``Finish'' button to finish the iteration, and then the searching result will be applied to their melody in the composing mode; Or, they can also click the ``Cancel'' button to discard all the iterations and go back.

\section{Algorithms and Implementations}

\subsection{Generative Model of Melodies}

Our implementation uses the pre-trained MusicVAE model \cite{Roberts:ICML:18} for melody generation, whose latent space is 512-dimensional.
Before we feed 2-bar melody to MusicVAE, we covert it from MIDI format to \emph{NoteSequence} representation \cite{NoteSequence}.
As it is a VAE architecture, it can encode a melody (which we denote by $\mathbf{x} \in \mathcal{X}$) to a latent vector (which we denote by $\mathbf{z} \in \mathcal{Z}$) as well as decode a latent vector to a melody.
We denote the encoder by $E: \mathcal{X} \to \mathcal{Z}$ and the decoder by $D: \mathcal{Z} \to \mathcal{X}$.
That is, given a latent vector $\mathbf{z}$, the decoder maps it to a melody $\mathbf{x}$; similarly, given a melody $\mathbf{x}$, the encoder maps it to a latent vector $\mathbf{z}$.
Note that any other deep generative models can be used as long as they support both encoding and decoding.

\subsection{Dimension Reduction of Latent Space}

It is known that BO does not work efficiently with very high-dimensional problems (e.g., more than 20 dimensions) \cite{Wang:JAIR:16,Chiu:SIGGRAPH:20}, and thus we need to apply dimensionality reduction to obtain a reasonably lower-dimensional latent space.
We tested several techniques for this purpose, including principle component analysis, auto-encoders, and VAEs.
We empirically found that VAEs worked well in terms of minimizing reconstruction errors and generating convincing and various melodies.
We also tested several different dimensionalities and found that reducing into a 4-dimensional latent space would be a reasonable choice.
Thus, we decided to train a VAE whose latent space is 4-dimensional.
Note that this choice is consistent with the previous work by Dinculescu et al.\ \citeyear{48628}.
For the training, we selected ten songs (resulting in 284 two-bar melody fragments) from the MusicVAE training dataset, the genres of which cover pop, rock, country, and R\&B.

In the following sections, we will still denote the (reduced) latent vector by $\mathbf{z}$ and the (reduced) latent space by $\mathcal{Z}$;
but notice that the dimension is 4, not 512.
Also, we will suppose that the encoder $E$ and decoder $D$ include the dimensionality conversion process.

Note that this dimensionality reduction would somewhat sacrifice the original expressiveness of the generative model;
we consider this is a limitation in our approach.
However, we found that the reduced model can still produce various melodies and is sufficiently inspiring and useful for helping novices find the desired melodies.

\subsection{Problem Formulation}

Following the previous work \cite{Brochu:NIPS:07, Koyama:SIGGRAPH:17, Koyama:SIGGRAPH:20}, we formulate the search of the user's desired melody as an optimization problem:
\begin{equation}
  \mathbf{z}^{*} = \mathop{\text{arg max}}_{\mathbf{z} \in \mathcal{Z}} f(D(\mathbf{z})),
\label{eqn:bo}
\end{equation}
where the function $f: \mathcal{X} \to \mathbb{R}$ represents the user's subjective preference over melodies, or how good the melody is (thus it is called a ``goodness function'' \cite{Koyama:SIGGRAPH:17}), and $\mathbf{z}^{*}$ is the latent vector that generates the most satisfying melody.
Note that we do not know the landscape of $f$, which is the challenge of this problem.
The goal here is to solve this optimization problem by iteratively presenting candidate solutions to the user and asking them to choose preferred ones.
In the current implementation, we assume a boundary $\mathcal{Z} = [-2.5, 2.5]^{4}$ to make the search problem tractable.

We generate candidate melodies by a preferential variant of BO \cite{Brochu:NIPS:07, Koyama:SIGGRAPH:17, Koyama:SIGGRAPH:20}.
In short, new candidates are determined based on all the user's feedback obtained so far so that every new candidate is as effective as possible (\autoref{sec:algorithm:acquisition}).
As in the previous work, we assume a Gaussian process (GP) prior on the goodness function.
We jointly estimate the underlying goodness values and the hyperparameters in GP by \emph{maximum a posteriori} (MAP) estimation;
refer to \cite{Koyama:SIGGRAPH:17} for details.

The algorithm takes the user's relative assessment as input.
In our system, the set of candidates involved at the $i$-th iteration consists of (1) $\mathbf{x}_{i}^{+}$, which is the current-best candidate (i.e., the user's selection at the last iteration) and (2) $\mathbf{x}_{i}^{(j)} = D(\mathbf{z}_{i}^{(j)}) \:\: (j = 1, 2, 3, 4)$, which are candidates sampled by latent-space interpolation between the current-best candidate and a BO-generated candidate (see \autoref{sec:algorithm:acquisition}), where the corresponding interpolation weights are 0.25, 0.5, 0.75, and 1.0.
We use Slerp for the interpolation following \cite{Roberts:ICML:18}.
In addition, (3) $\mathbf{x}_{i}^\text{edited}$, which is the candidate after the user's editing, is also involved here.
We interpret this situation as relative-comparison feedback in latent space as
\begin{equation}
  E(\mathbf{x}_{i}^\text{edited}) \succ \{ E(\mathbf{x}_{i}^{+}), \mathbf{z}_{i}^{(1)}, \mathbf{z}_{i}^{(2)}, \mathbf{z}_{i}^{(3)}, \mathbf{z}_{i}^{(4)} \},
\end{equation}
which means that the latent vector $E(\mathbf{x}_{i}^\text{edited})$ is preferred over the other latent vectors on the right-hand side.
This preferential feedback can be probabilistically modeled by the Bradley--Terry--Luce (BTL) model \cite{Tsukida:TR:11}.

\subsection{Candidate Selection}
\label{sec:algorithm:acquisition}

In BO methods, the next candidates are determined by maximizing an \emph{acquisition function}, which calculates the effectiveness of a candidate as the next query.
There exist several acquisition functions \cite{Shahriari:ProcIEEE:16}, but the most popular choice is probably the one called the \emph{expected improvement} (EI), as used in \cite{Brochu:NIPS:07, Koyama:SIGGRAPH:17, Koyama:SIGGRAPH:20, Chong2019InteractiveSE}.
It is defined in our problem setting as
\begin{equation}
  a^\text{EI}(\mathbf{z}) = \mathbb{E} [ \max\{ 0, f(D(\mathbf{z})) - f^{+} \} ],
\end{equation}
where $f^{+}$ is the estimated goodness value of the current-best melody.
The maximization of the EI function is efficiently solved by standard optimization algorithms (e.g., L-BFGS).

\section{Algorithm Evaluation}

We conducted an experiment from the algorithmic viewpoint, where we artificially simulated user behaviors.
The goal is to verify whether our algorithm can find the desired melody in a reasonable number of iterations.
We designed this experiment based on the assumption that there is one specific melody, which is the user's desired melody.
In the experiment, we used \emph{soft dynamic time warping} (s-DTW), an improvement of the original DTW, as the quantitative similarity metric between two melodies in latent space \cite{pmlr-v70-cuturi17a}.

We simulated user behaviors as follows.
First, we generated a desired melody, $\mathbf{x}^\text{target}$, by random sampling on the latent space.
Given the set of candidate melodies in the $t$-th iteration, $\{ \mathbf{x}_{t}^{+}, \mathbf{x}_{t}^{(1)}, \ldots, \mathbf{x}_{t}^{(4)} \}$, we selected the one with the smallest s-DTW to $\mathbf{x}^\text{target}$ as the best preferred one at the iteration.
We iterated this procedure for $t = 1, \ldots, 50$.
We performed 100 trials in total.
In addition to the s-DTW values, we also recorded the 4-dimensional latent vectors of all the selected candidate melodies and calculated the Euclidean distances to the latent vector of the desired melody.
Note that we regenerated the desired melody for every trial.

\begin{figure}
  \centering
  \includegraphics[width=\textwidth]{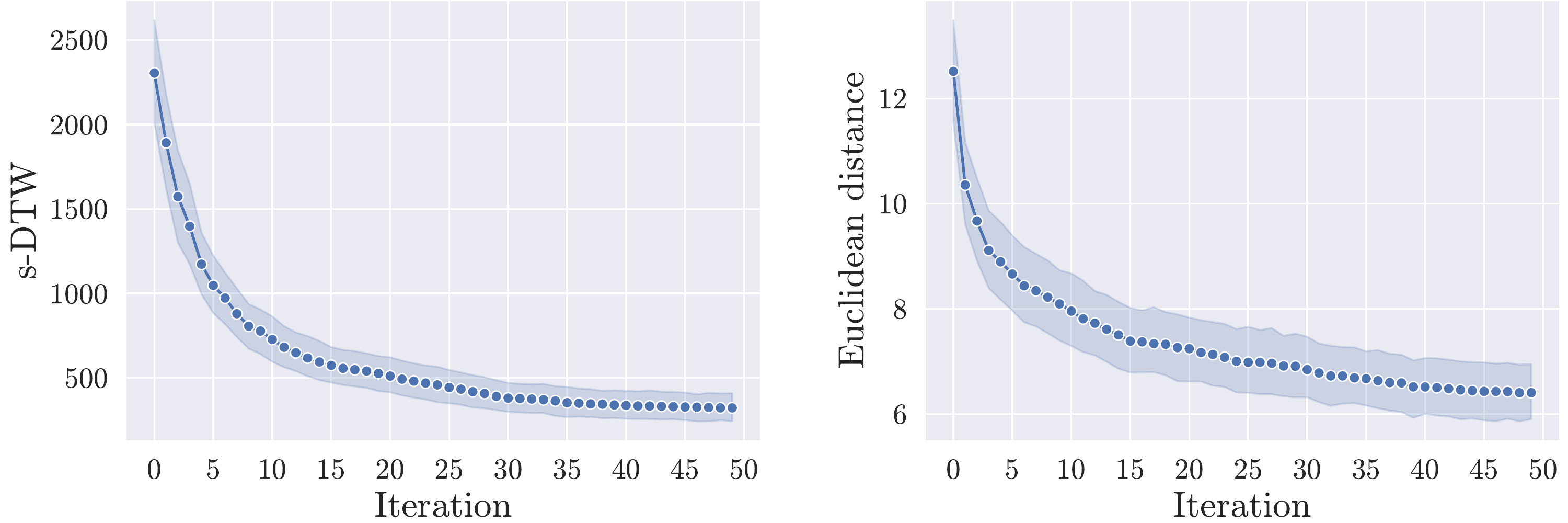}
  \caption{
    Simulation experiment results. Each plot shows the mean of 100 trials with the colored region showing the standard deviation. The left shows s-DTW and the right shows the Euclidean distance in the 4-dimensional latent space.
  }
  \label{fig:result}
\end{figure}

\autoref{fig:result} (left) shows the plot of the s-DTW values, and \autoref{fig:result} (right) shows the plot of the corresponding latent-space Euclidean distances.
Each dot on the plot represents the mean of 100 trials.
We could observe that the s-DTW values decreased as iterations went on, which was our expectation.
We found the Euclidean distances in the latent space also decreased consistently, which indicates the smoothness of the latent space.
The results suggest that BO could effectively find melodies that are reasonably similar to the target one after around 20 iterations in the 4-dimensional latent space.

\section{Pilot Study}

To obtain initial qualitative feedback on the proposed approach, we conducted an informal pilot study with two novice composers, who had wished to compose music but had no formal composition training.
We asked one of the two participants ($P_1$) to compose a 32-bar melody, and the other one ($P_2$) to compose a 16-bar melody with our system.
Each participant was first tutored by the experimenter about the features of the system.
During the experiment, the participants were free to compose melodies by themselves or use the ``Search'' function to get help from the system.
When they used the ``Search'' function, they were asked to go for at least 20 iterations.

The composed melodies can be found at the our web source\footref{note1}.
$P_1$ finished the composing within 2.5 hours, while $P_2$ finished it within 1.5 hours.
In the experiment, $P_1$ used the ``Search'' function 8 times, while $P_2$ used it 6 times.

In the follow-up interview, both the participants explicitly expressed that the system had effectively provided them more ideas than they had, which inspired their creativity.
\emph{``Before I used the `Search' function, I didn't realize that there could be so many possibilities developed from my idea''} ($P_1$).
\emph{``After I heard the candidates suggested, I got some fantastic ideas based on them''} ($P_2$).
Both of the participants also mentioned that the system made the process of composition easier.
\emph{``I tried to compose before but got stuck in extending my idea. But in this experiment, just `Search'. It saves my life''} ($P_1$).
\emph{``I liked it because it seemed that those four candidates in each iteration were sorted from similar to different, as compared to my original idea. It made it easier to select one to fit my melody''} ($P_2$).

However, they also mentioned room for improvement or drawbacks they noticed.
First, $P_1$ mentioned that she wanted more control on the search function, saying \emph{``It would be better if I can set the degree of how different, or how aggressive the candidates to be. Because I already had some ideas on how the melody goes''}.
Besides, $P_2$ wanted the system to take the sequential musical information into account when searching for candidates, saying \emph{``Even though it found some variations on the melody clip I composed in a previous paragraph, sometimes the candidates found did not fit a new paragraph''}.

\section{Conclusion}

In this paper, we demonstrated that human-in-the-loop BO had the potential to work effectively and efficiently on searching for the user's desired melody from the latent melody space.
We also showed that BO could suggest satisfying candidates within a reasonable number of iterations.
The pilot study result suggested that it also inspired the user's creativity and lowered the learning barriers of novice composers.
In the future, we plan to investigate how to apply BO to a higher dimensionality (e.g., 512) efficiently and effectively in an interactive system.
We also want to provide more control in terms of the searching behavior, and to consider the sequential information, as indicated by the pilot study participants.
Finally, we hope our work can inspire researchers to more deeply investigate designing music creation tools, or even creativity support tools in a broader context, by incorporating the human-AI mixed-initiative interaction mechanism.

\bibliographystyle{apacite}
\bibliography{CSMC_MUME_LaTeX_Template.bib}

\end{document}